\documentclass[%
 aip,
 amsmath,amssymb,
 reprint,%
]{revtex4-1}
\usepackage[dvipsnames]{xcolor}
\usepackage{graphicx}
\usepackage{dcolumn}
\usepackage{bm}

\usepackage[utf8]{inputenc}
\usepackage[T1]{fontenc}
\usepackage{mathptmx}
\usepackage{etoolbox}

\makeatletter
\def\@email#1#2{%
 \endgroup
 \patchcmd{\titleblock@produce}
  {\frontmatter@RRAPformat}
  {\frontmatter@RRAPformat{\produce@RRAP{*#1\href{mailto:#2}{#2}}}\frontmatter@RRAPformat}
  {}{}
}%
\makeatother
\begin{document}

\preprint{AIP/123-QED}

\title[High-Power Test of a C-band Linear Accelerating Structure with an RFSoC-based LLRF System]{High-Power Test of a C-band Linear Accelerating Structure with an RFSoC-based LLRF System}
\author{C. Liu}

 \email{chaoliu@slac.stanford.edu}
 
\author{L. Ruckman}%
\author{R. Herbst}%
\author{D. Palmer}%
\author{V. Borzenets}%
\author{A. Dhar}%
 \affiliation{ 
SLAC National Accelerator Laboratory, Menlo Park, California, USA.
}%

\author{D. Amirari}%
\author{R. Agustsson}%
\author{R. Berry}%
 \affiliation{ 
RadiaBeam Technologies LLC, Santa Monica, California, USA.
}%

\author{E. Nanni}%
 \affiliation{ 
SLAC National Accelerator Laboratory, Menlo Park, California, USA.
}%

\date{\today, The following article has been accepted by Review of Scientific Instruments by the AIP.}

\begin{abstract}
Normal conducting linear particle accelerators consist of multiple rf stations with accelerating structure cavities. Low-level rf (LLRF) systems are employed to set the phase and amplitude of the field in the accelerating structure and to compensate the pulse-to-pulse fluctuation of the rf field in the accelerating structures with a feedback loop. The LLRF systems are typically implemented with analogue rf mixers, heterodyne based architectures and discrete data converters. There are multiple rf signals from each of the rf stations, so the number of rf channels required increases rapidly with multiple rf stations. With a large number of rf channels, the footprint, component cost, and system complexity of the LLRF hardware will increase significantly. To meet the design goals of being compact and affordable for future accelerators, we have designed the next generation LLRF (NG-LLRF) with a higher integration level based on RFSoC technology. The NG-LLRF system samples rf signals directly and performs rf mixing digitally. The NG-LLRF has been characterized in loopback mode to evaluate the performance of the system and has also been tested with a standing-wave accelerating structure, a prototype for the Cool Copper Collider (C\(^3\)) with peak rf power level up to 16.45 MW. The loopback test demonstrated amplitude fluctuation below 0.15\% and phase fluctuation below 0.15 degree, which are considerably better than the requirements of C\(^3\). The rf signals from the different stages of accelerating structure at different power levels are measured by the NG-LLRF, which will be critical references for the control algorithm designs. The NG-LLRF also offers flexibility in waveform modulation, so we have used rf pulses with various modulation schemes which could be useful for controlling some of rf stations in accelerators. In this paper, the high-power test results at different stages of the test setup will be summarized, analyzed and discussed.
\end{abstract}

\maketitle

\section{Introduction}

The low-level radio frequency (LLRF) control systems for future particle accelerators could be extremely challenging with stringent field stability requirements and hardware size, weight and power consumption (SWaP) and cost specifications. The conventional LLRF system detects the amplitude and phase of the rf signal from the accelerating structure by down converting the rf signals to intermediate frequency (IF) by analogue rf mixers and then sampling by analogue to digital converter (ADC). The feedback control algorithm implemented in field programmable gate array (FPGA) or processors uses the digital samples from the ADC to calculate the updated base-band pulse waveform to compensate for the fluctuations in rf field. The new base-band pulse waveform is generated by a digital to analogue converter and then up converted to rf by vector modulators \cite{geng2017rf}. For control and monitoring purposes, there are multiple rf signals from the different stages for each rf station, such as, pre-amplifier output, klystron output, klystron reflected and rf signals from several couplers in each accelerating structure. With a large number of rf stations, the footprint and cost of the LLRF system increases significantly with the large number for rf channels required. For future accelerators with more stringent requirement in size and cost, the hardware design of LLRF could become challenging. We have designed and implemented the next generation LLRF (NG-LLRF) system for cool copper collider (C\(^3\)), which has been proposed as a compact and affordable system for lepton collider Higgs factory \cite{bane2018advanced,dasu2022strategy,vernieri2023cool,nanni2023status}. The NG-LLRF system is implemented based on the AMD Xilinx RFSoC technology, which integrates high-speed data converters, field programmable logic (FPGA) and processors in a single device. The high integration level of the NG-LLRF offers a more compact and cost-effective LLRF platform solution for future particle accelerators.

We have developed control and readout platforms for a range of high energy physics (HEP) and astrophysics instruments that operate in the GHz frequency range \cite{liu2021characterizing, liu2022development,  liu2023evaluating, henderson2022advanced, liu2023higher,liu2024development}. Based on the lessons learned for RFSoC devices, the NG-LLRF prototype system has been developed. The initial prototype platform with a custom solid state amplifier (SSA) demonstrated a phase jitter as low as 87.54 fs \cite{liu2024direct}. In \cite{liu2024generationllrfcontrolplatform}, the hardware, firmware and software of the NG-LLRF were described and the initial high-power test results demonstrated that the NG-LLRF can drive and measure the rf system with a high precision. 

In this paper, the high-power test of the C\(^3\) accelerating structure with NG-LLRF performed at the Radiabeam C-band high-power test facility will be summarized and discussed. The NG-LLRF has been used as the rf drive of the test setup and the measurement platform for all the rf signals at different stages of the test setup. The high-power test setup and the test procedure will be introduced in Section \ref{setup}. Then the pulse-to-pulse stability evaluation results for phase and amplitude at different drive stages will analyzed in Section \ref{pulse_to_pulse}. For high-power test results, the measurements with rf signal modulated with square wave at different peak power levels and pulse widths will be discussed in Section \ref{square_pulse}. As the rf modulating processing of NG-LLRF is fully implemented in the digital domain, it offers high flexibility in modulating the rf pulse with different waveform shapes. Phase modulation schemes are used for multiple purposes at different types of rf stations in the accelerators. The phase reversal schemes have been commonly used for rf pulse compressor since it was first implemented for SLAC Energy Doubler (SLED) \cite{farkas1974sled,lin2022x}. In Section \ref{phase_reversal}, the test results of high-power rf pulse with phase reversal will be demonstrated and discussed. Compensating the beam loading effect is one of the main objectives for the LLRF systems of accelerators and phase modulation has been simulated and experimented for beam loading compensation \cite{kashiwagi1998beam,mastoridis2017cavity}. In Section \ref{phase_sweep}, the rf signals from the accelerating structure driven by a linear phase sweep rf pulse will be analyzed to understand the phase modulation behavior of the high-power test circuit.

\section{High-Power Test Setup} \label{setup}
\begin{figure}
  \begin{center}
  \includegraphics[width=3in]{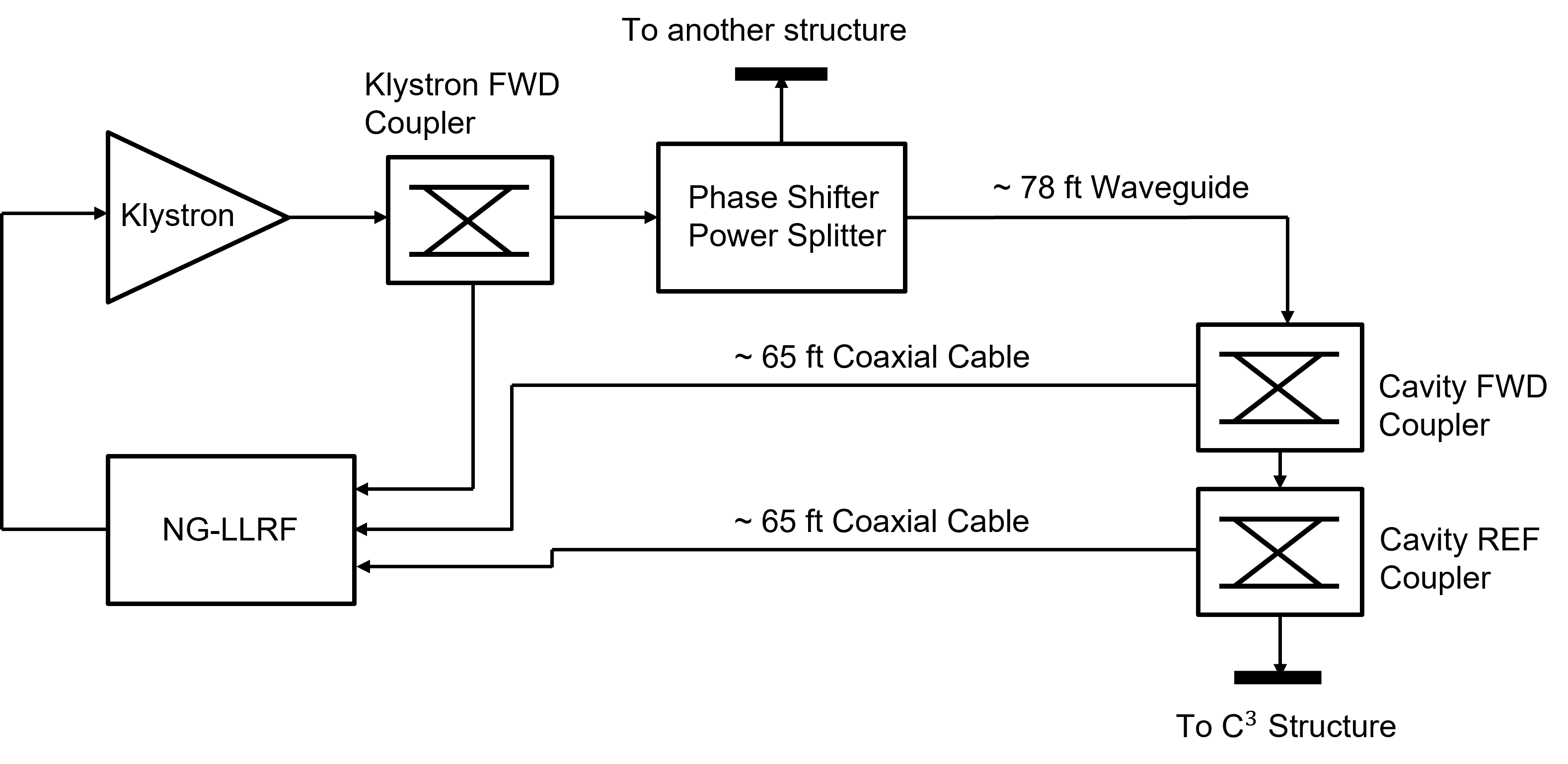}\\
  \caption{The waveguide schematics of the high-power test setup. The NG-LLRF platform generates the rf pulses to drive the test setup and measures the rf signals at different stages of the setup via directional rf couplers.}\label{fig-1a}
  \end{center}
\end{figure}

\begin{figure}
  \begin{center}
  \includegraphics[width=3.4in]{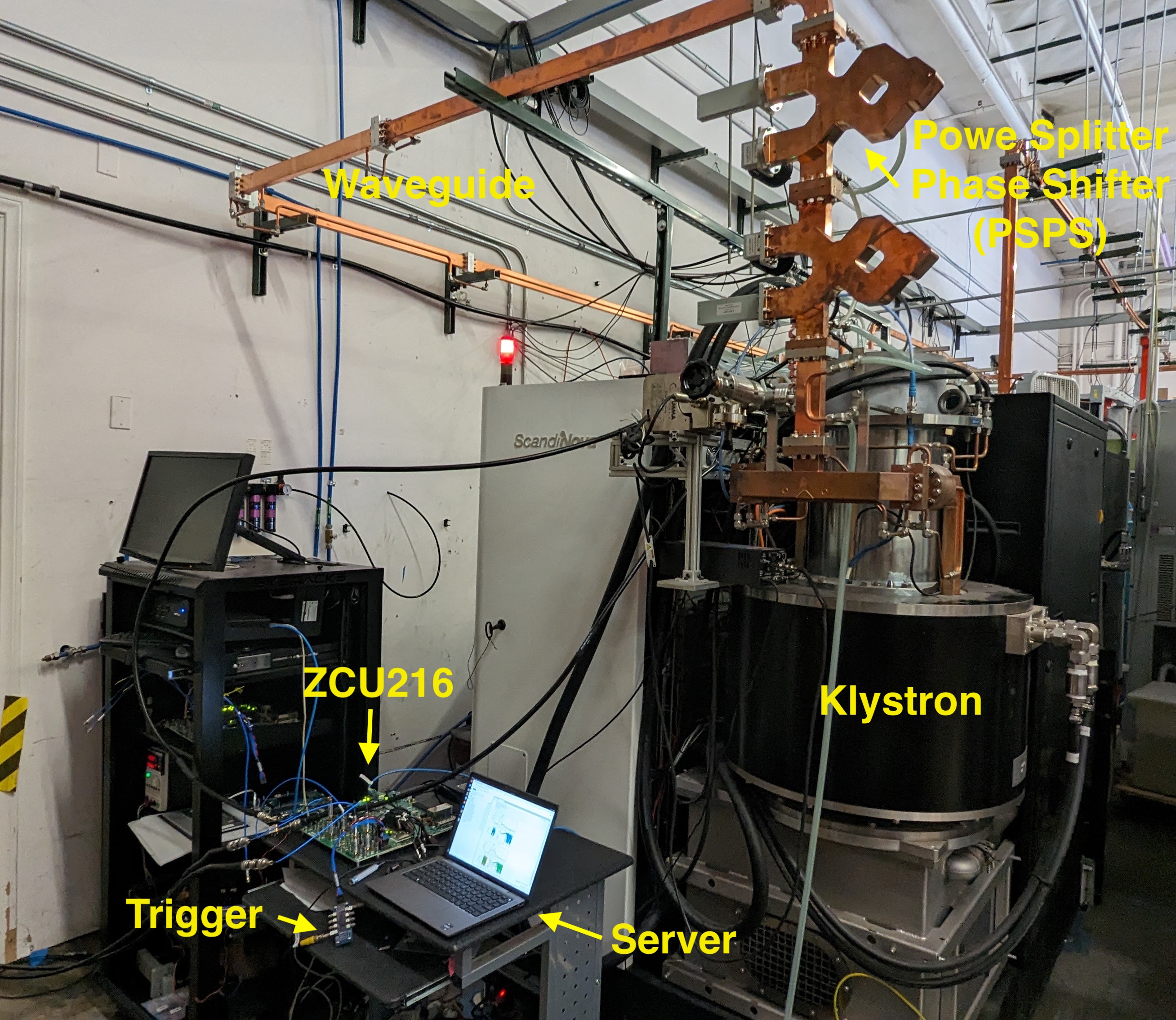}\\
  \caption{The rf drive system of the high-power test setup.  The rf pulse generated by the NG-LLRF is amplified by a solid state amplifier (SSA) and drives the klystron. The power splitter and phase shifter (PSPS) divides the klystron output rf power to two separate test bunkers.}\label{fig-1}
  \end{center}
\end{figure}
The high-power test facility at Radiabeam, including the C-band klystron, waveguide and the test bunker, has been used for the tests summarized in this paper. The waveguide schematics of the high-power test setup is shown in Figure \ref{fig-1a}. Figure \ref{fig-1} highlights the main components of the drive system. The NG-LLRF generates the rf pulse with frequency around 5.712 GHz. The rf pulse is amplified by an integrated solid state amplifier (SSA) and then injected to the klystron. The maximum peak power output of the klystron is approximately 50 MW. The klystron drives two independent accelerator structures and the C\(^3\) structure is one of them. The output rf power of the klystron is divided by a power splitter and phase shifter (PSPS) and around 47\% of the total power is delivered to the C\(^3\) structure via the WR-187 waveguide, which is a rectangular waveguide with a long side length of 1.872 inches.
\begin{figure}
  \begin{center}
  \includegraphics[width=3in]{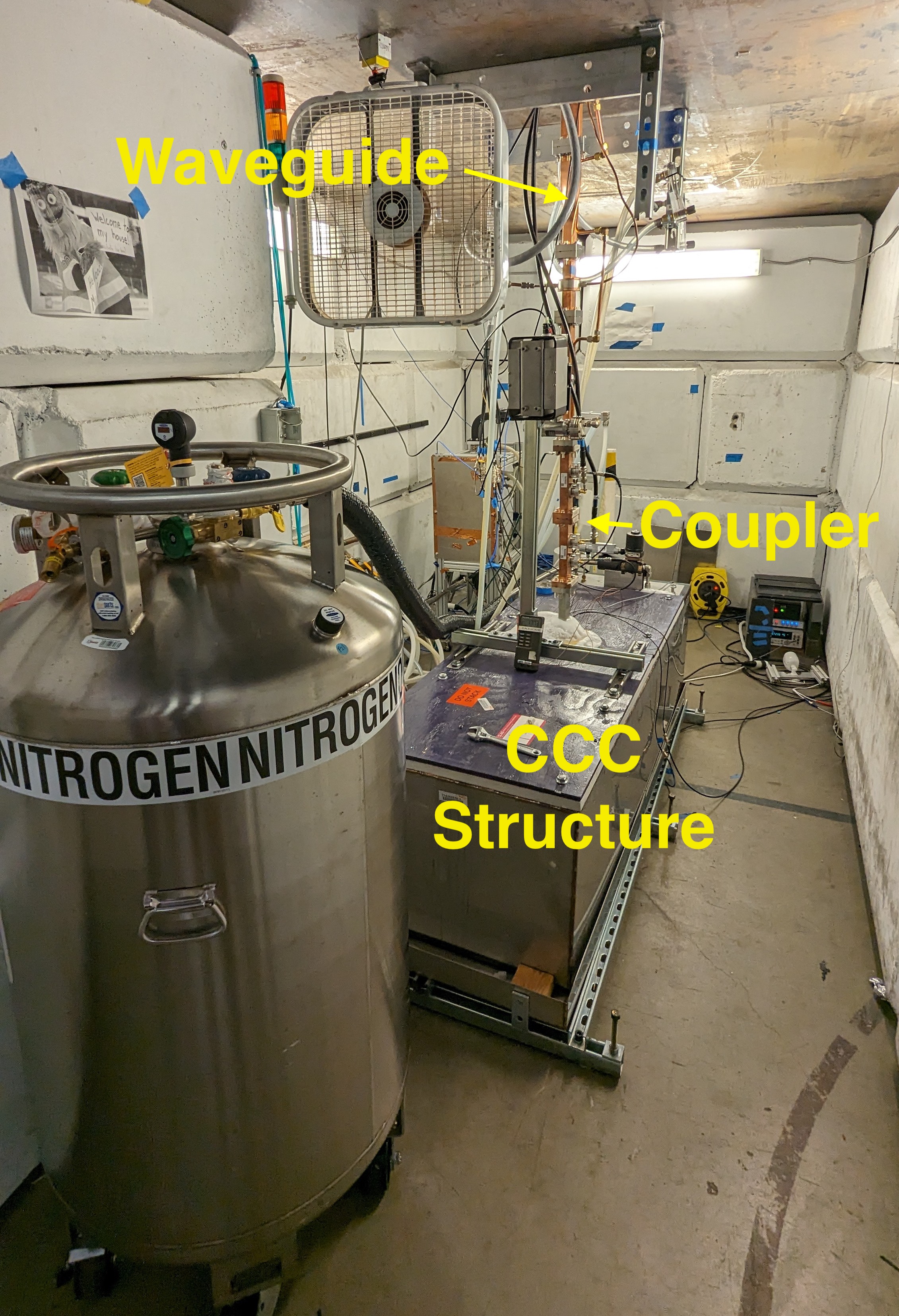}\\
  \caption{The setup in the test bunker for the high-power test. The rf power is delivered to the C\(^3\) accelerating structure. The rf signals from two coupler ports are looped back to to the NG-LLRF via 60 ft long coaxial cables.}\label{fig-2}
  \end{center}
\end{figure}

\begin{figure}
  \begin{center}
  \includegraphics[width=3in]{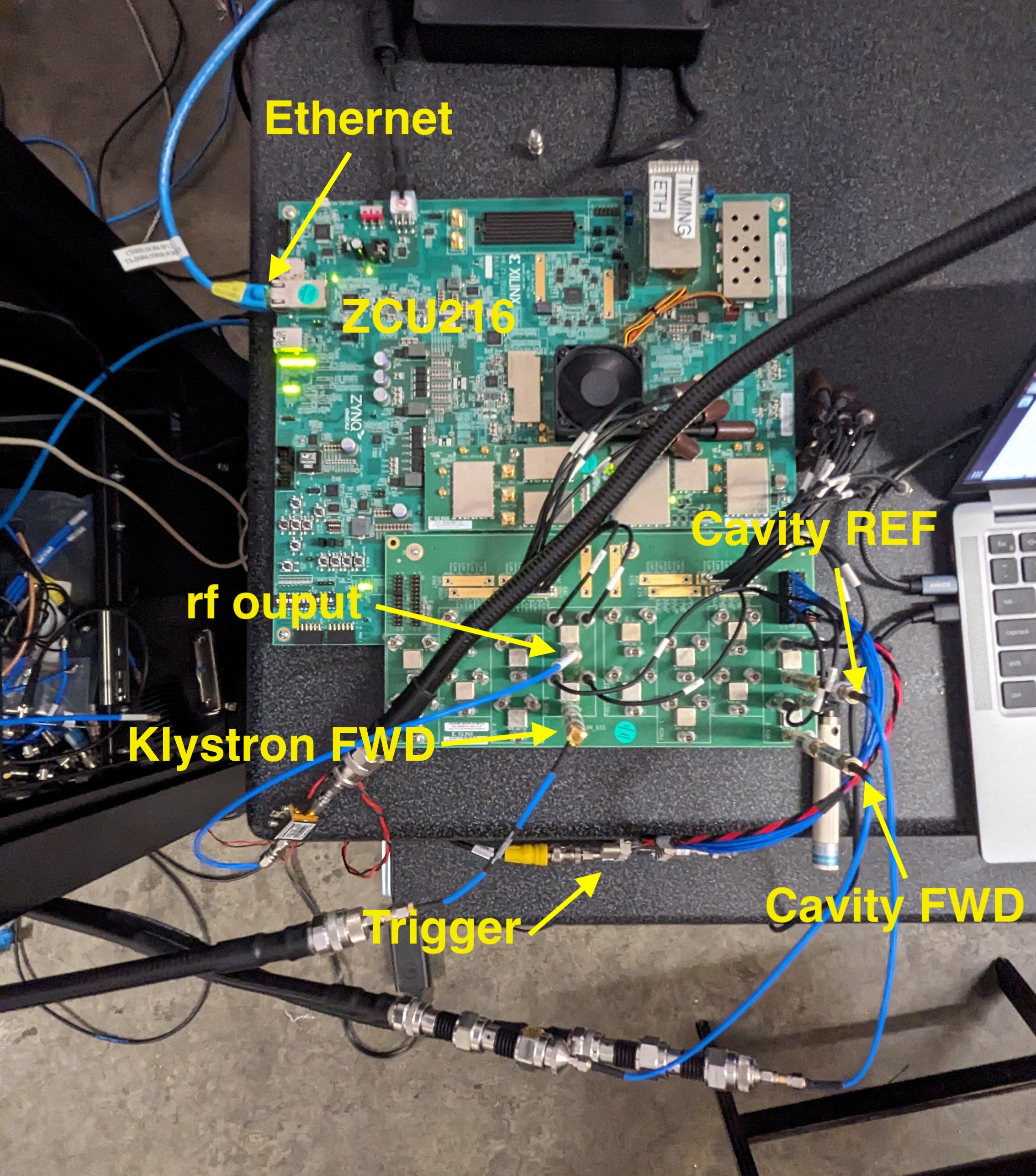}\\
  \caption{The hardware of the prototype NG-LLRF sytem. The rf output generates the rf pulse that drives the entire high-power test setup. The rf signals from different stages of the test setup are sampled by the rf inputs of the NG-LLRF. Both the rf pulse generation and rf signal capturing is triggered by an external source.}\label{fig-3}
  \end{center}
\end{figure}

Figure \ref{fig-2} shows the test setup in the bunker. The rf power is delivered to the C\(^3\) accelerating structure operating at a liquid nitrogen temperature of 77 K in the tank shown in Figure \ref{fig-2}. The peak rf power injected to the C\(^3\) structure in this test ranges from  4.2 to 16.45 MW. Between the waveguide and the C\(^3\) accelerating structure, there are two directional couplers to sample the rf signals from the structure, one for the forward and the other for the reflection (also see Figure \ref{fig-1a}). The coupling loss for forward and reflection couplers are -64.3 dB and -65.6 dB respectively.  The rf signals from the couplers are attenuated by 40 dB attenuators and the cable loss is approximately 4.2 dB. The maximum power looped back to the NG-LLRF is below the 1 dBm maximum input power of the integrated ADCs in RFSoC. The attenuated cavity forward (FWD) and cavity (REF) are connected to two of the rf input channels of the NG-LLRF prototype hardware built based on a Xilinx ZCU216 evaluation board as shown in Figure \ref{fig-3}. In this experiment, we also measure the klystron output via a directional coupler with one of the rf input channels, which is the klystron forward (FWD) labeled in Figure \ref{fig-3}. The rf output generates the rf pulse that drives the test setup. The rf pulse generation and rf signal measurements are triggered by an external trigger source, which also triggers the SSA and klystron.   

\section{Pulse-to-pulse Stability Analysis} \label{pulse_to_pulse}

Stabilizing the amplitude and the phase of the field in the accelerating structure is the main objective of a LLRF system. Therefore, the phase and amplitude pulse to pulse fluctuation levels are critical for LLRF systems. In both \cite{liu2024direct} and \cite{liu2024generationllrfcontrolplatform}, the phase and amplitude fluctuation results with SSAs have been summarized and discussed. In this test, the stability levels at multiple stages of the high-power test setup will evaluated. The fluctuation or noise contributions from each stage will be analyzed and discussed. 
\begin{figure}
  \begin{center}
  \includegraphics[width=3.4in]{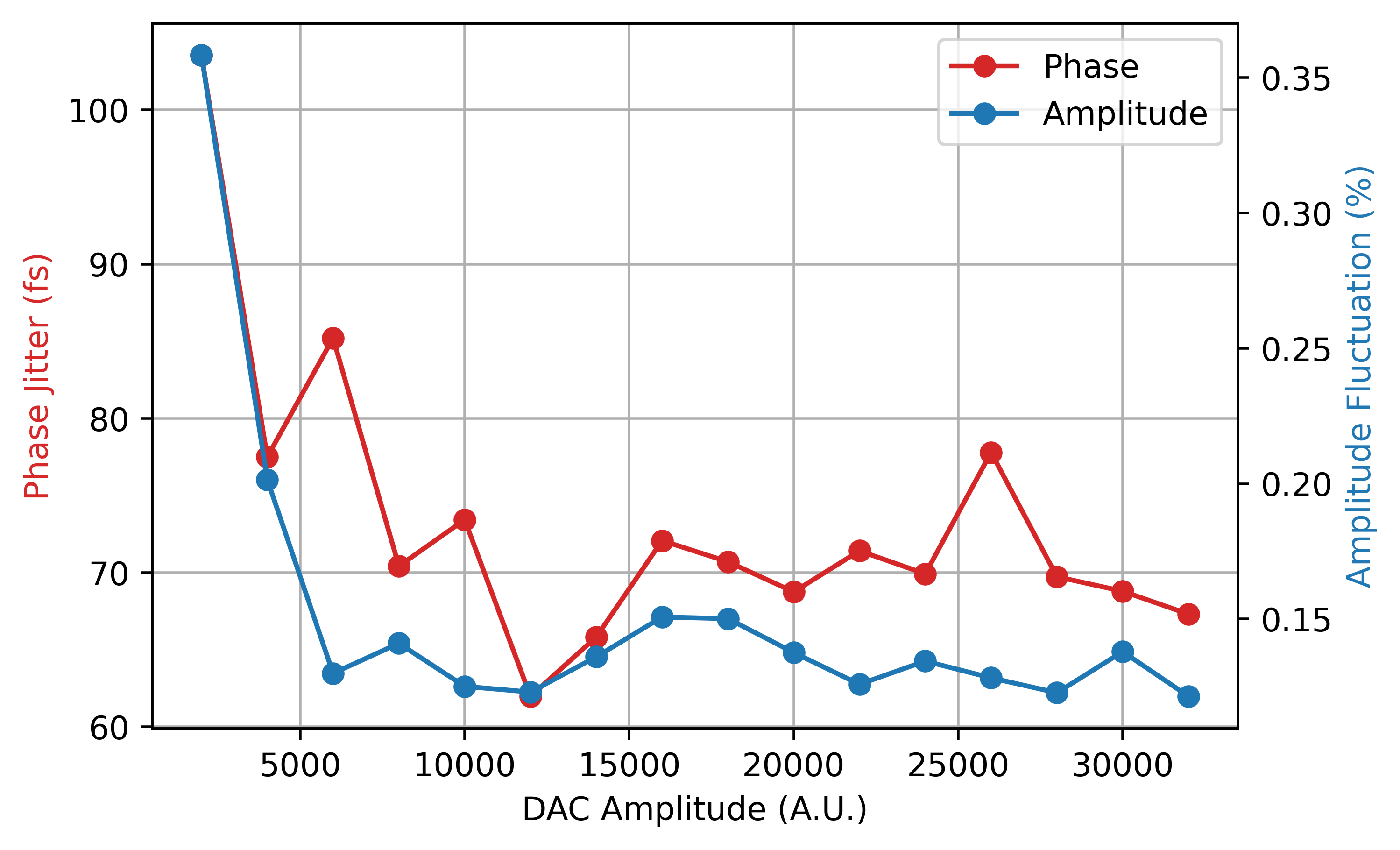}\\
  \caption{The phase and amplitude stability of the NG-LLRF in direct loop-back mode with DAC configured to generate rf pulse with different amplitude levels. The phase jitter is converted to time units and the amplitude fluctuation is measured in percentage with respect to the average amplitude in each of the measurements.}\label{fig-4}
  \end{center}
\end{figure}

\begin{figure}
  \begin{center}
  \includegraphics[width=3.4in]{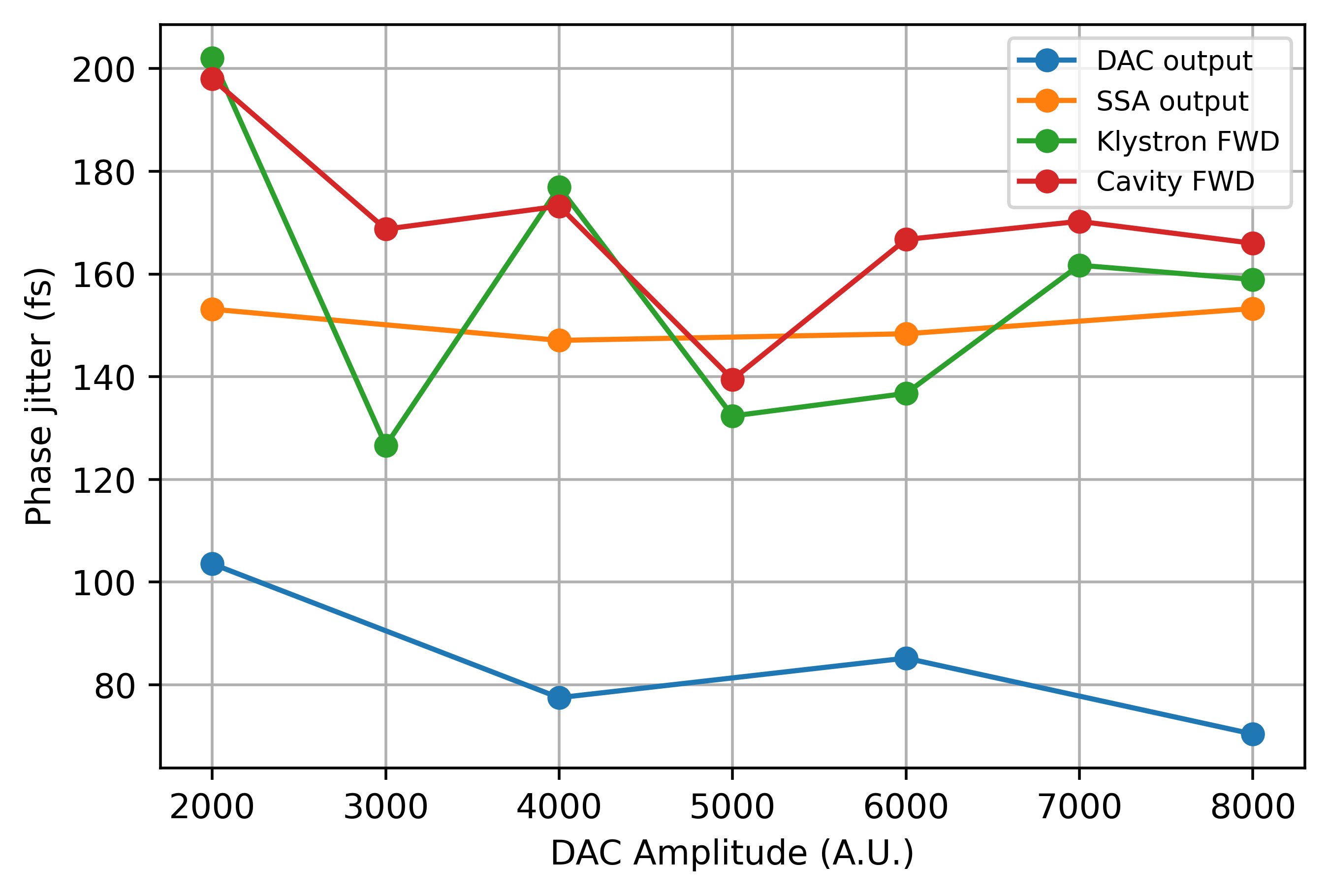}\\
  \caption{The phase jitter levels at different stage of the driving circuit at different power levels.}\label{fig-5}
  \end{center}
\end{figure}

The first test is the direct loop-back from the rf output DAC to one of the rf input ADCs via a passive band-pass filter. This is a critical test for the NG-LLRF accelerator applications as it determines the highest stability level can be achieved by the platform. The DAC generates the rf pulse at 5.712 GHz and 1 \(\mu\)s duration at 60 Hz. In the digital domain, the rf pulses are generated by modulating an rf signal with a square pulse and the test has been performed with square pulses at different amplitude levels. The rf pulses are decoded and generated by the integrated DAC in RFSoC, so the amplitude levels are labeled as DAC amplitude on the horizontal axis in Figure \ref{fig-4} and \ref{fig-5}. For each amplitude level, 60 consecutive pulses are captured and processed offline in software to evaluate the stability. The down-converted data samples have been captured in the format of in-phase (I) and quadrature (Q) pairs. The I and Q components are converted to magnitude and phase values for performance analysis. The average phase and amplitude values on the flat-top on each of the 60 pulses are calculated. The phase fluctuation is measured with the standard deviation of the 60 average phase values. The amplitude fluctuation is measured in percentage of the standard deviation of the 60 average amplitude values with respect to the average of all the average amplitude values of the pulses. 

Figure \ref{fig-4} shows the phase and amplitude fluctuation levels at DAC amplitude from 2000 to 32,000 in 2000 steps and 32,000 is close to the maximum amplitude for modulation. The phase fluctuation levels are converted from the unit of degree to the corresponding phase jitters in time units. The phase jitter drops significantly from over 100 fs at 2000 DAC amplitude to around 70 fs with DAC amplitude higher than 8000. A phase jitter of 70 fs at 5.712 GHz is equivalent to 0.14\textdegree. The phase jitter of the loop-back setup are considerably better than the 150 fs requirement of C\(^3\) \cite{nanni2023status}. The amplitude fluctuation follows the same trend as the phase and fluctuates at around 0.13\%, which is close to the 0.1\% requirement of C\(^3\) \cite{nanni2023status}. The NG-LLRF demonstrates high stability when the DAC amplitude is 8000 or above. To achieve the optimum stability performance, at least 25\% of the full ranges of DACs and ADCs should be used. In conclusion, the NG-LLRF data converters with direct rf sampling can drive and measure the C-band rf signals with adequate precision to achieve the field stability requirement. 

The phase and amplitude fluctuation levels at different stages of the high-power test setup have been evaluated by a similar method with the direct loop-back test. In this case, the rf pulse generated by the NG-LLRF drives the entire setup and the output signals of SSA, klystron and rf signal from cavity forward are measured by the rf input channels of the NG-LLRF. The test has been performed while the accelerating structure conditioning \footnote{With the field intensity required for accelerator applications, structures require time to condition up to high field values. During the conditioning period, plasma breakdowns are observed as the surfaces of the structure and high-power rf components are conditioned. To avoid damages to the test setup, short rf pulses are utilized to limit the stored energy.} is still in progress, so the pulse width has been reduced to approximately 450 ns to achieve the peak power of 16.45 MW at 60 Hz. The phase jitter levels at different stages are also characterized with 60 consecutive pulses. As the orange trace in Figure \ref{fig-5} shows, the SSA added around 80 fs of additional phase jitter to the rf pulse compared with the direct loop-back test. The phase jitter levels of klystron forward fluctuate around the SSA jitter level, which is approximately 150 fs. The phase jitter added by the field coupling from the klystron to the accelerating structure is in the range of 10 to 40 fs. The measurements at high power level are more robust, as both the NG-LLRF and the klystron are operated in desired ranges. With 16.45 MW delivered to the accelerating structure, the added phase jitter by the SSA, klystron and waveguide are around 82.8 , 5.6 and 7.0 fs respectively, which totals the final phase jitter to 166 fs. With a feedback control, the 150 fs jitter requirement of C\(^3\) is highly achievable with the NG-LLRF platform.The development of a real-time feedback control loop is still in progress, and more test results will be published with a complete implementation of the LLRF system.

\section{High-power tests with different rf pulse shapes}

Amplitude and phase modulation techniques are commonly used for pulse compression, beam loading compensation and other purposes in different rf stations of particle accelerators. As the modulation and demodulation of the NG-LLRF are fully implemented in digital domain, the rf pulse can be modulated with any pulse shapes. In this section, the rf signals from the C\(^3\) accelerating structure driven by rf pulses with different shapes will be discussed.

\subsection{Square rf pulse} \label{square_pulse}

The high-power test begins with the setup driven by a rf signal modulating with square pulses. The rf signals from the C\(^3\) accelerating structure have been captured at different peak power levels delivered to the structure. The peak power delivered to the structure ranges from 4.2 to 16.45 MW with different pulse widths and pulse rates. The higher power levels and pulse rates have been achieved with a 450 ns pulse width, but we are focusing on 1\(\mu\)s pulse width in this section. With pulse width of 1\(\mu\)s, the highest peak power that can be delivered to the structure during this phase of conditioning was approximately 5.2 MW at 10 Hz within the desired breakdown rate of 100 per second. 

\begin{figure}
  \begin{center}
  \includegraphics[width=3.4in]{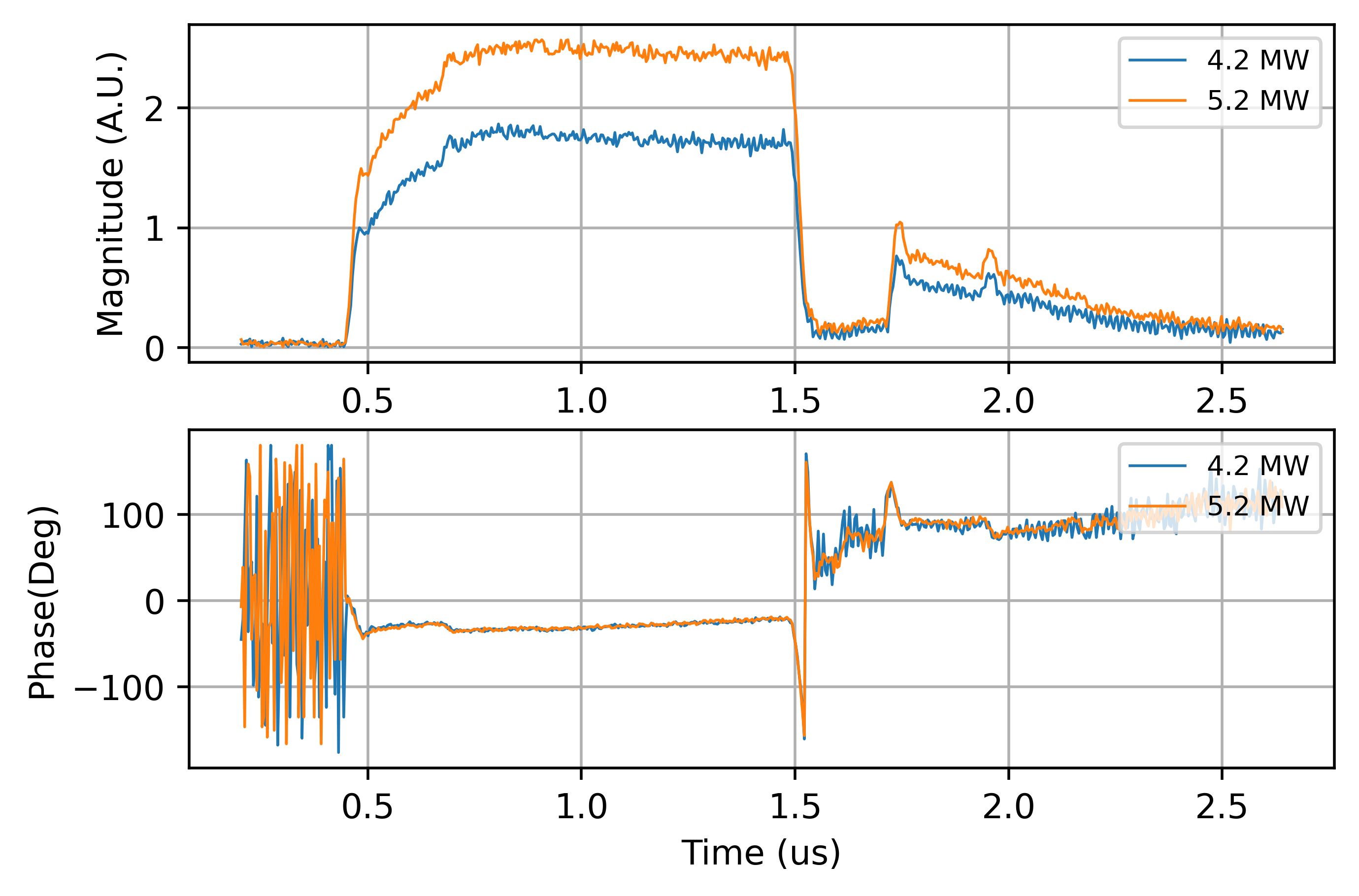}\\
  \caption{The klystron forward signals captured with 1 \(\mu\)s rf pulses with peak power of 4.2 MW and 5.2 MW. }\label{fig-6}
  \end{center}
\end{figure}

\begin{figure}
  \begin{center}
  \includegraphics[width=3.4in]{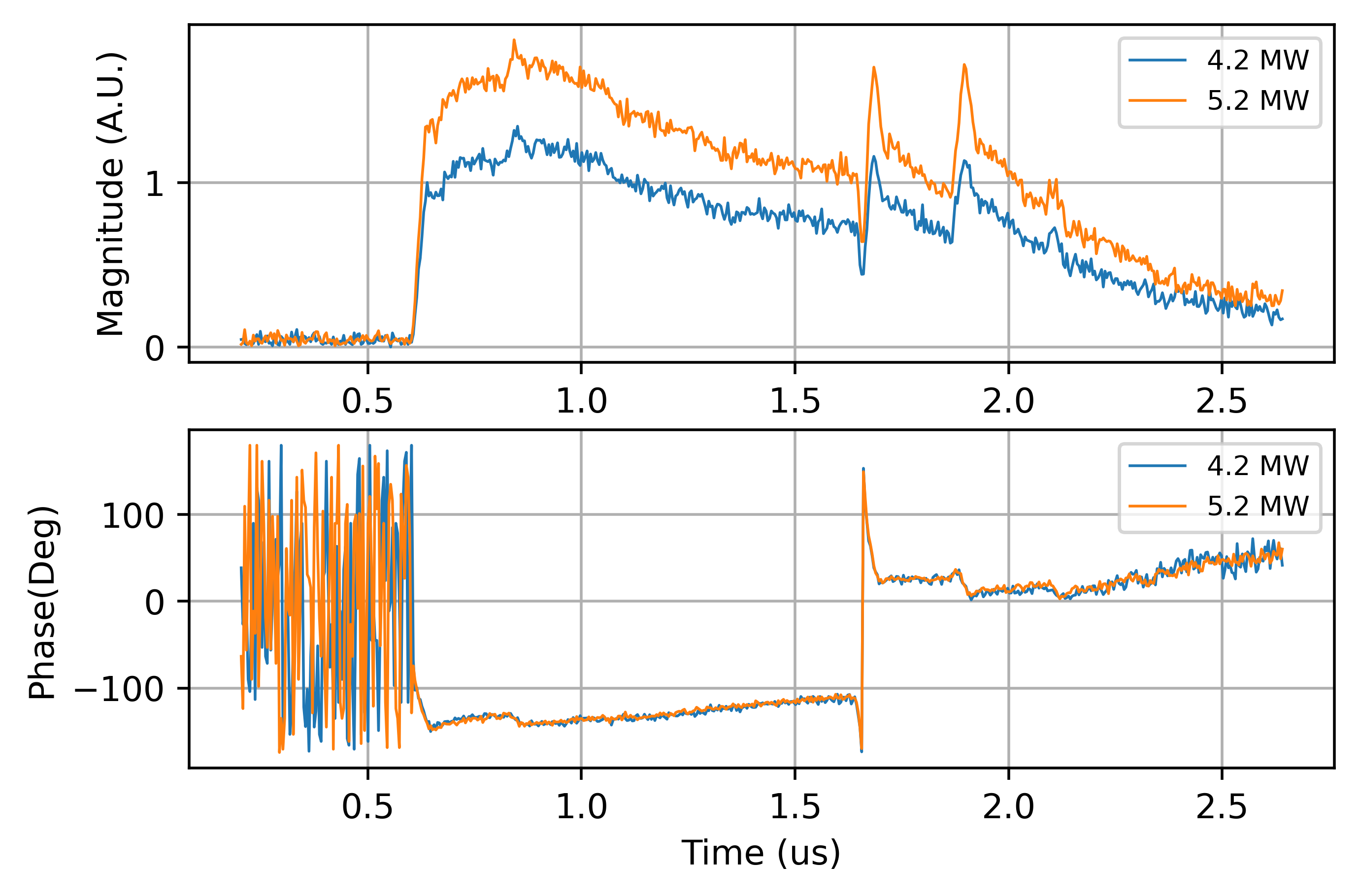}\\
  \caption{The cavity forward signals captured with 1 \(\mu\)s rf pulses with peak power of 4.2 MW and 5.2 MW. }\label{fig-7}
  \end{center}
\end{figure}

The klystron forward and cavity forward rf signals from the structure driven by a 1\(\mu\)s and 10 Hz at peak power of 4.2 and 5.4 MW are shown in Figure \ref{fig-6} and Figure \ref{fig-7}. The magnitudes of rf signals in both figures have been normalized by the first peak of the signal captured at lower power. As the figures show, the klystron forward and cavity forward signals at different power levels have similar trends in magnitude and extremely close phase values, which demonstrates the high consistency and precision of rf pulse generation and measurement of the NG-LLRF prototype. 

As Figure \ref{fig-6} shows, the klsytron forward signal ramps up to a flat-top in approximately 0.5 \(\mu\)s, which follows the rise time of SSA as shown in \cite{liu2024generationllrfcontrolplatform}. After reaching the peak, the klystron forward signal begins to decrease slightly. As the klystron operates significantly below saturation, it is sensitive to the voltage droop in the modulator, which results in loss of gain. Therefore, the magnitude of the forward signals declines with a slight slop until the rf power is down. The klystron forward signal drops sharply when rf is off. However, the signal rises after around 350 ns, which is the rf power reflected back from the C\(^3\) accelerating structure.

\begin{figure}
  \begin{center}
  \includegraphics[width=3.4in]{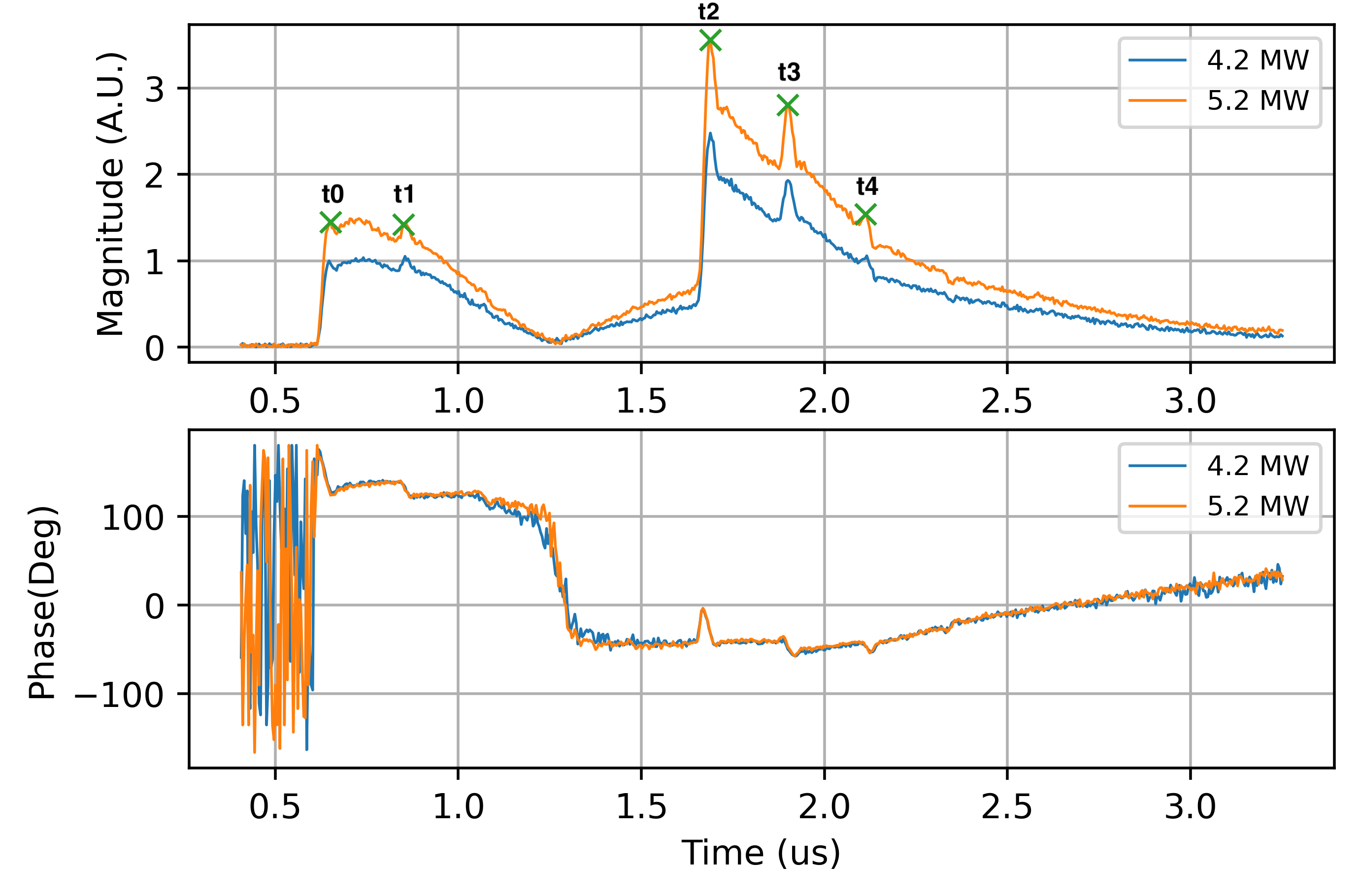}\\
  \caption{The cavity reflection signals captured with 1 \(\mu\)s rf pulses with peak power of 4.2 MW and 5.2 MW. }\label{fig-8}
  \end{center}
\end{figure}

The magnitude of the cavity forward signal shown in Figure \ref{fig-7} ramps up as the rf power propagates to the structure and declines with a noticeable slope until the rf power is down. This slope is due to the cross coupling between the forward and reflected signals at the directional coupler. This cross talk can also be confirmed by observing the signal after the rf drive is turned off. In the ideal case, the cavity forward signal would fall rapidly after the rf power is switched off. However, the reflected power can also couple to the forward coupler, which appears as the rf signal spikes when the rf drive is turned off (see Figure \ref{fig-7}) and then ramps down slowly with second and third peaks in the magnitude traces. If we observe the reflected signal in Figure \ref{fig-8}, we see that there is a close correlation in this time range. Given that the response matches the reflected signal in time and that the directional couplers are adjacent, we can determine the directionality of the signal. A true measurement of the ``FWD" amplitude would be determined around 1.25 \(\mu\)s, as the reflected signal is 0 at this moment in time.

\begin{figure}
  \begin{center}
  \includegraphics[width=3.4in]{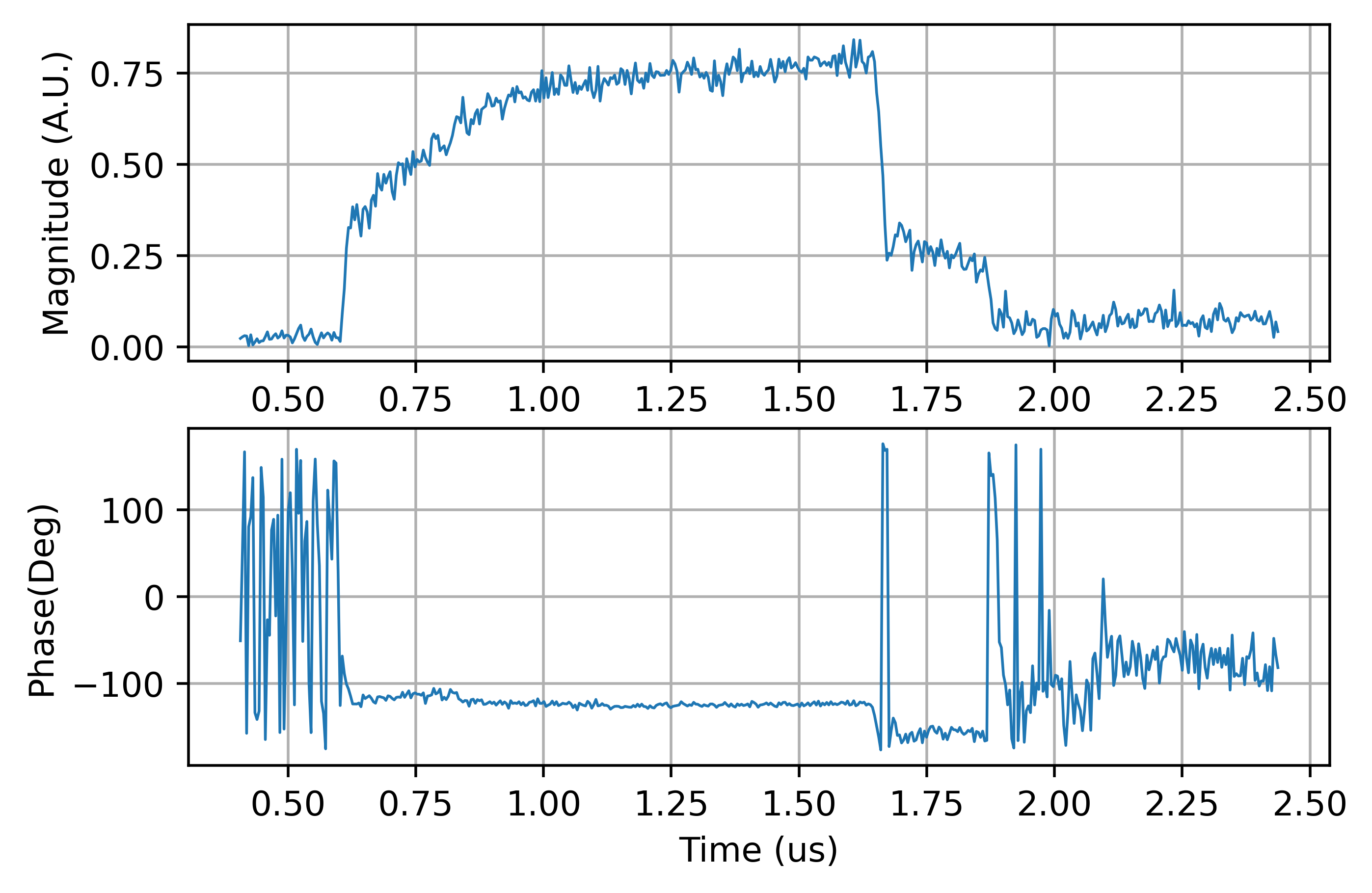}\\
  \caption{The cavity forward signal subtracted by cavity reflection signal. Both of the rf signals are captured with 1 \(\mu\)s rf pulses with peak power of 5.2 MW. }\label{fig-8a}
  \end{center}
\end{figure}

Figure \ref{fig-8} shows the cavity reflection signals at peak power levels of 4.2 and 5.2 MW. The rf field is filling the accelerating structure from t0 and the magnitude of reflection signals drops linearly with time. There is a spike at t1 on the linear ramp, which is around 199.6 ns after t0. The spike at t1 corresponds to the reflected power from the structure reflects back to the PSPS and then get reflected to the structure and reflected back again to the cavity reflection coupler. The magnitude of the reflection signal at two power levels reaches zero in approximately 550 ns. After this time, the reflected rf signal from the structure increases in magnitude and inverts in phase indicating that the cavity is overcoupled. This is consistent with the low power rf measurements that were performed on the structure after installation. When the rf power is switched off, another peak occurs at t2 on the cavity reflection signal, which is 133\% higher than the first peak. The accelerating structure cavities have been designed to be critically coupled with the beam. As there is no beam present, the accelerating structure is overcoupled. The second peak and the decay followed correspond to the discharging process of the structure cavities. The rf power reflected from the C\(^3\) prototype structure flows back to the waveguide in the reversed direction and then the power is reflected back by the PSPS to the structure and reflected again by the structure, which appears as the second peak as the signals decay at t3 on the signal sampled with the cavity reflection coupler. The same reflection mechanism repeats again and results in the third peak on the decay at t4. The time gap between t2 and t3 is around 211.6 ns, which is almost exactly the same as the time gap between t3 and t4. 

The time gaps between t0 and t1, t2 and t3 and t3 and t4 should be the same as the paths of reflection signals are identical. However, due to the rise time of the klystron, the gap between t0 and t1 is approximately 12 ns shorter than the other two gaps. With the full length of the waveguide and the group velocity of the waveguide, the time gaps in theory can be calculated.

For the WR-187 waveguide employed by this test, the group velocity \(v_{gr}\) can be calculated by

\begin{equation}\label{vgr}
    v_{gr} = v\sqrt{1 - \left(\frac{f_c}{f}\right)^2}= v\sqrt{1 - \left(\frac{\lambda}{\lambda_c}\right)^2}
\end{equation}

In Equation \ref{vgr}, \(v\) is the speed of light in linear medium. And the \(f_{c}\) is the cutoff frequency of waveguide and \(f\) is the operating rf frequency, which is around 5.712 GHz in this case. The \(\lambda\) is the wavelength at the operating rf frequency. Half of wavelength at the cutoff frequency is equivalent to the long side length of waveguide \(a\), so the wavelength at the cutoff frequency can be expressed as

\begin{equation}\label{lambda}
    \lambda_c = 2a
\end{equation}

For the test setup, the \(v_{gr}\) calculated is 83.38\% of speed of light through a vacuum space. In the time gaps discussed above, the reflection rf waves propagate in the full waveguide twice before being captured by the cavity reflection coupler again. As Figure \ref{fig-1a} shows, the length of the waveguide between the PSPS and the test structure is 78 ft and the adapters, couplers and other structures added another 5.3 ft waveguide length. The total time gaps between the peaks calculated is 203.1 ns, which only 8.5 ns less than the time gap measured between t2 and t3 or t3 and t4 in Figure \ref{fig-8}. The minor difference could be primarily attributed to the errors on measuring the total length of the waveguide. In general, the calculation and measurement results are highly matched. 

To further analyze the cross talk between the forward and reflection couplers, the reflection signal has been subtracted from forward signal and the the residual signal is shown in Figure \ref{fig-8a}. In this case, the rf signals captured with peak power of 5.2 MW are employed for analysis. The rf signals are normalized to the peak at t3 in Figure \ref{fig-8} and reflection signal has been phased shifted to match the phase of the forward signal at t3. The residual signal shown in Figure \ref{fig-8a} has similar profile with the klystron forward signals shown in Figure \ref{fig-6} before the rf pulse has been switched off, which has flat magnitude and phase levels after 1 \(\mu\)s. The peaks and the decay trend on the cavity forward signal after the rf pulse switched off shown in Figure \ref{fig-7} have been almost eliminated by the subtraction. The profile of the residual signal further verified the existence of the cross coupling and revealed the true shape of the rf pulse that has been injected into the accelerating structure. The cross coupling can be primarily attributed to a mismatched load on the cavity forward coupler, which could be optimized in the future high power test setup. 

\begin{figure}
  \begin{center}
  \includegraphics[width=3.4in]{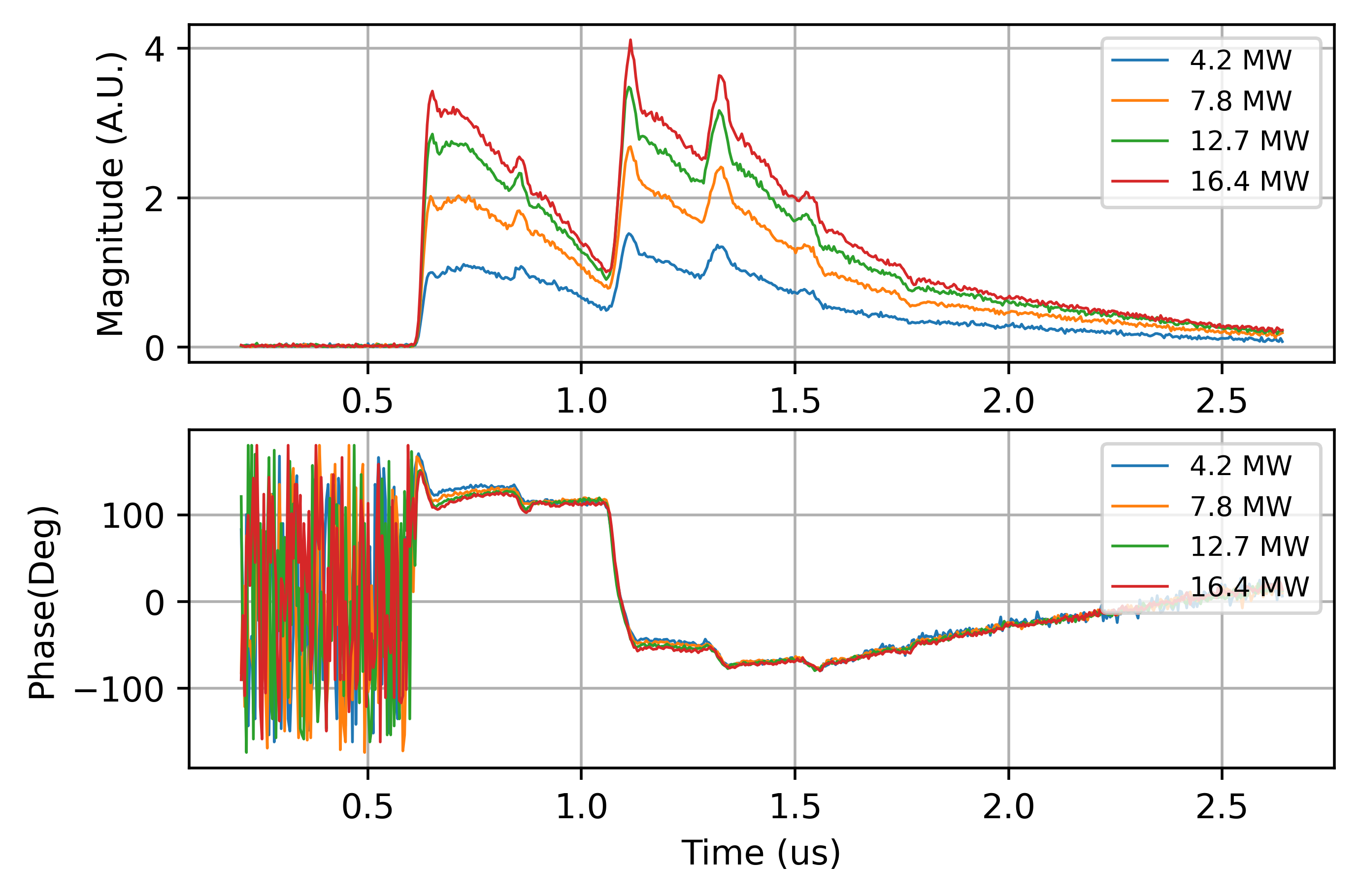}\\
  \caption{The cavity reflection signals captured with 450 ns rf pulses with peak power at different levels. }\label{fig-7b}
  \end{center}
\end{figure}

The high-power test has also been performed with higher peak power levels at 60 Hz. To operate the setup within desired breakdown rate,  the pulse duration has been reduced to approximately 450 ns. Figure \ref{fig-7b} shows the the cavity reflection pulse captured at different peak power levels from 4.2 to 16.45 MW. The reflection signals are scaled to the first peak of the measurement at 4.2 MW peak power. As the power level increases, the gradient of the declines in the field filling process increases significantly. Due to the shorter pulse duration, the rf field filling has not been completed before the rf power is down. The second peak corresponding to the overcoupling effect is higher than the first peak also occurs after the rf power is off. As the filling time is shorter, the second peak is only around 50\% higher than the first peak. The decay corresponds to the dissipation process has similar trends for all the power levels.

For future linear colliders like C\(^3\), the bunch trains with small bunch spaces are employed to achieve the desired luminosity \cite{nanni2023status}. With single-bunch operation, the fluctuation within each pulse is less critical than pulse-to-pulse fluctuation. However, for the multi-bunch operation, the fluctuation level within each of the pulses becomes critical. As the cavity forward rf signals shown in Figure \ref{fig-7}, there is a noticeable slope on the magnitude traces, which is the result of cross coupling between the cavity forward and reflection couplers. After the cross coupling is minimized by subtracting the reflection signal from forward signal, the residual signal can represent the "true" cavity forward signal, which has a 550 ns flat-top before the rf switches off. The flat-top needs to be extended to 700 ns to meet the requirement of C\(^3\). After the structure is fully conditioned, the klystron can be operated closer to saturation and deliver rf power pulse with higher stability within the pulse. In the future high power test with a fully conditioned structure and the klystron operating at optimum power level, more tests with rf pulses with longer duration and higher repetition rate will be performed to further demonstrate the stability of the NG-LLRF platform. The beam loading effect on the bunch train can also introduce significant fluctuations to the forward power \cite{kashiwagi1998beam}, so an algorithm to compensate for the fluctuation within the pulse should be considered. The implementation of such algorithms will be greatly enhanced with the programmable signals and precision measurements of the NG-LLRF. In case of accelerators operated with long bunch trains, the beam loading effect could be sensed from the reflected rf signal from the accelerating structure. An algorithm can be developed to calculate the update amplitude and phase of the rf pulse based on the reflection signal measured by the NG-LLRF to compensate for the effect. As the beam loading effect should be pulse-by-pulse consistent, the rf pulse shape only needs to be tuned after the beam is switched on.

\subsection{Rf pulse with phase reversal}\label{phase_reversal}

Phase reversal applied to the rf pulse became common for pulse compressors since SLED has been implemented and it delivered stable performance. From the first SLED implementation to more recent modified versions, the phase reversals were realized with a separate phase shifter or inverter components. For the NG-LLRF platform, the modulation processes are fully implemented in digital domain. A base-band waveform in any shape can be generated in software and modulated in the digital data-paths integrated in RFSoC. In this case, the rf signal has been modulated with a square wave with a phase reversal every 200 ns to understand the behavior of the system with phase reversal. The rf pulse delivered to the test structure in this case has a peak power of 4.2 MW and a pulse duration of approximately 1 \(\mu\)s at 10 Hz.  

\begin{figure}
  \begin{center}
  \includegraphics[width=3.4in]{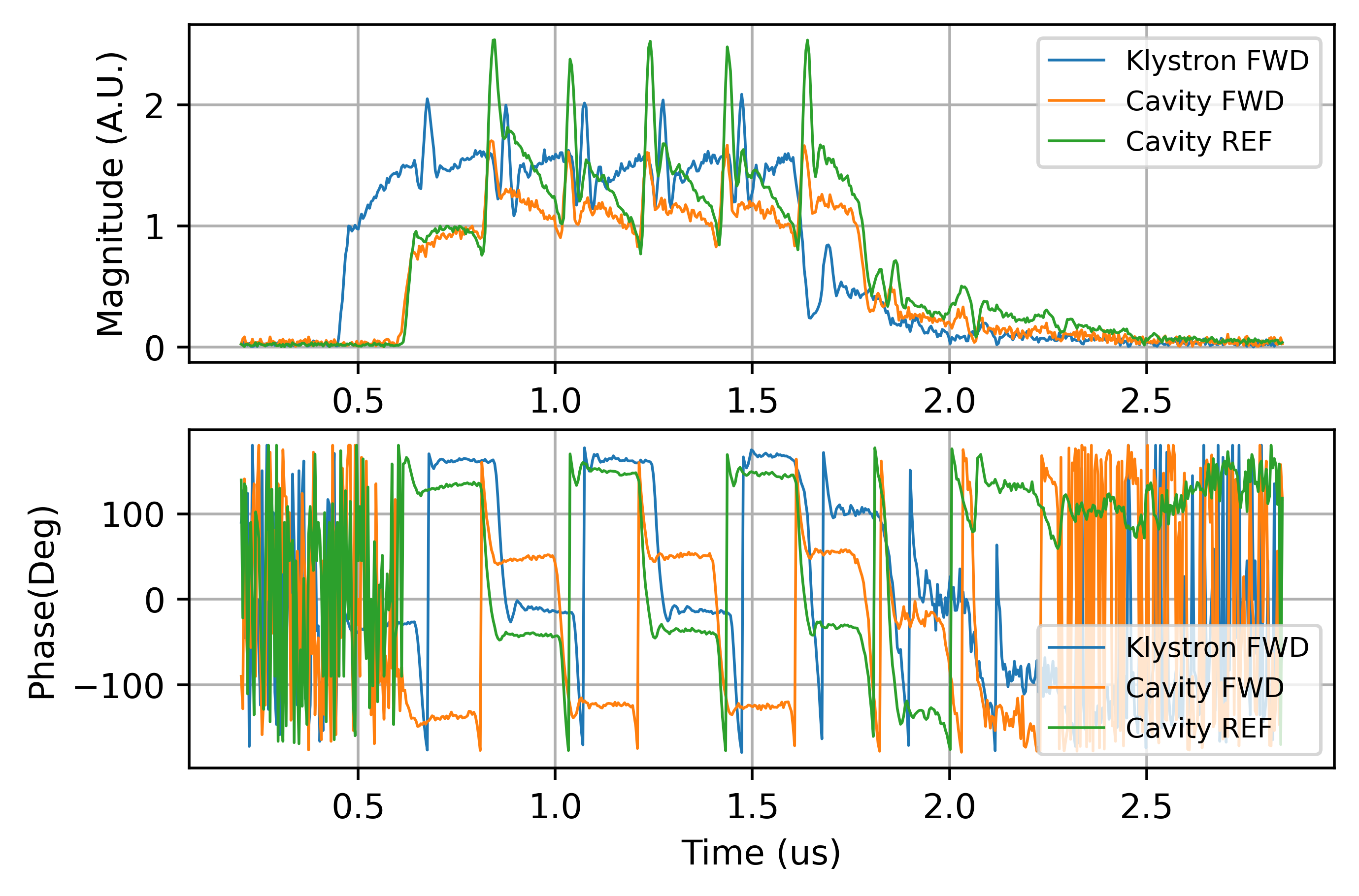}\\
  \caption{The klystron forward, cavity forward and reflection signals captured with 1 \(\mu\)s rf pulses with peak power of 4.2 MW. The rf pulse is modulated with a square wave with phase reversal every 200 ns. }\label{fig-9}
  \end{center}
\end{figure}

Figure \ref{fig-9} shows the rf pulses captured at the three different stages for the high-power test setup, which are klystron forward, cavity forward and cavity reflection. The magnitude of all three signals are normalized to the first peak on the rising edge of each pulse. The magnitude of the klystron forward pulse spikes around 30\% of the average value each time the phase of drive is reversed. The phase of the klystron forward pulse reversed by 180\textdegree \, every 200 ns with a spike follow by ripples settled in less than 100 ns. The cavity forward signal follows the similar waveform as the klystron forward signal. 

In the first 200 ns after the rf power arrived the structure, the cavity has been filled in the same way as it was driven by a square wave. When the phase is reversed at 200 ns, the rf power has been extracted by the field out of phase rapidly. The extraction appears as the second peak followed by a rapid fall on the cavity reflection signal, which is over 160\% higher than the first peak. This filling and extraction process cycle is similar to the method used by the SLED system to produce a short pulse with higher power.  After the field is completely extracted, the structure is filled with the rf field with reversed phase, which occurs as the slower decay after the rapid fall. When the phase is reversed again, the power extraction and filling repeat again, and that corresponds to the third peak and the decay followed. This cycle alternates until the rf power is down.

\begin{figure}
  \begin{center}
  \includegraphics[width=3.4in]{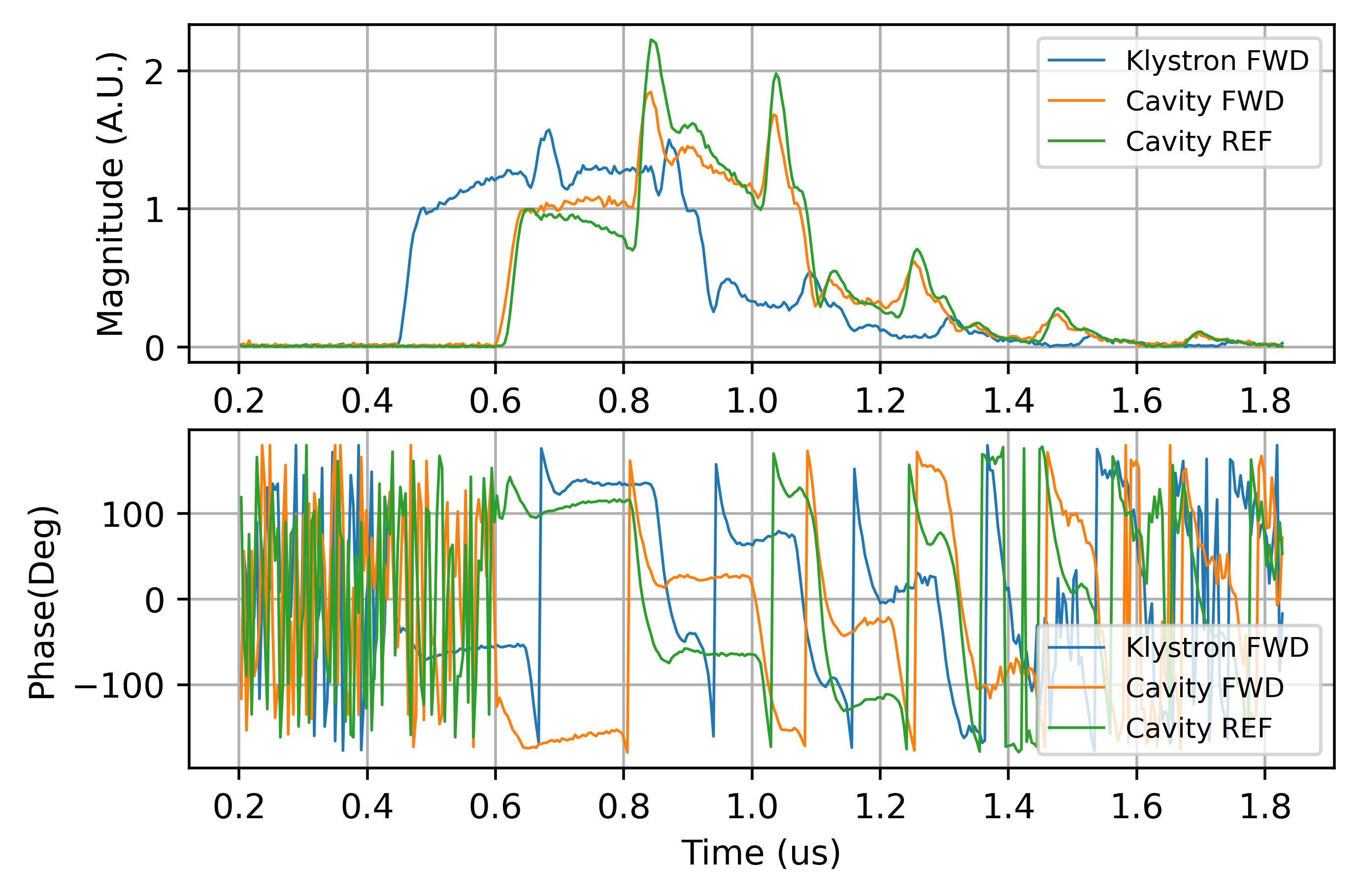}\\
  \caption{The klystron forward, cavity forward and reflection signals captured with approximately 450 ns rf pulses with peak power of 16.45 MW. The rf pulse is modulated with a square wave with 180 degree phase reversal every 200 ns. }\label{fig-11}
  \end{center}
\end{figure}

The phase reversal test has been performed with 450 ns rf pulses with peak power of 16.45 MW at 60 Hz. As Figure \ref{fig-11} shows, the rf signals captured at 16.45 MW follow trends similar to the signals measured at 4.2 MW shown in Figure \ref{fig-9}. As the power level is significantly higher, the cavity reflection signal declines with higher gradient and the field filling process is more obvious in the first 200 ns. In the power extraction and filling process after the second peak of the cavity reflection signal, the change in the gradient shows the transition between rapid field extraction and filling with an inverted phase.

\subsection{Rf pulse with linear phase sweep}\label{phase_sweep}

The phase modulation technique has been used to reduce the rf power requirement and beam loading compensation of High-Luminosity Large Hadron Collider (HL-LHC) \cite{mastoridis2017cavity}. The technique could be used for future particle accelerators for those purposes or beyond. In this case, the rf signal has been modulated with a linear phase ramp from -180\textdegree \, to 180\textdegree \, in 500 ns to understand the responses of the rf drive and the C\(^3\) structure to the phase ramp at peak power up to 16.45 MW at 60 Hz of pulse rate.

\begin{figure}
  \begin{center}
  \includegraphics[width=3.4in]{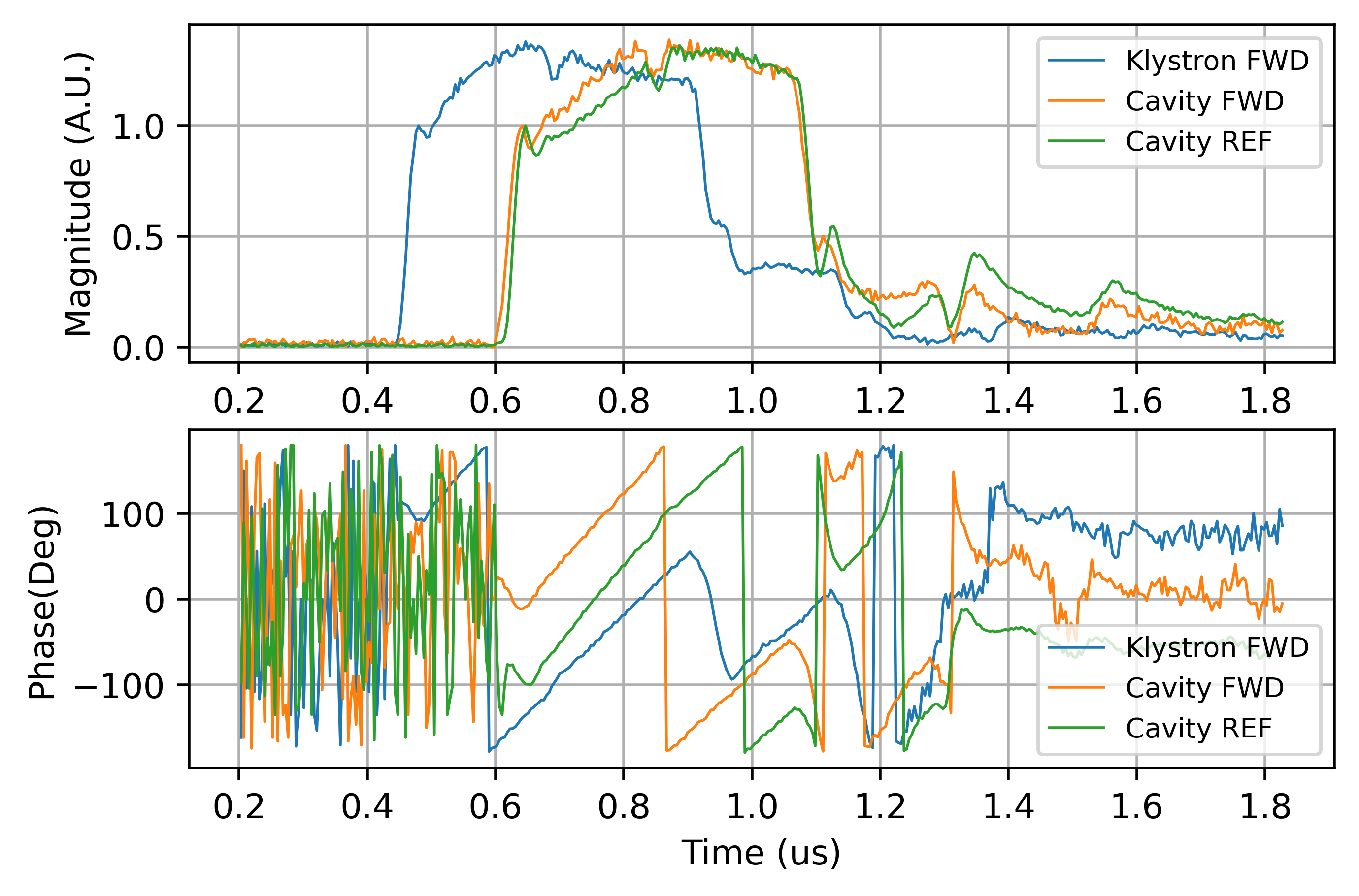}\\
  \caption{The klystron forward, cavity forward an reflection signals captured with approximately 500 ns rf pulses with peak power of 16.45 MW and a linear phase sweep within the pulse duration. }\label{fig-12}
  \end{center}
\end{figure}

Figure \ref{fig-12} shows the rf signals measured with rf pulse with 360\textdegree \, linear phase ramp at peak power level of 16.45 MW. The three rf signals have different initial phase values, but the 360\textdegree \, phase ramps are fully completed for all of them. The 360\textdegree \, phase linear sweep in 500 ns can also be considered as driving the test setup with around 2 MHz frequency offset. The klystron has the bandwidth to generate the phase sweep. However, the accelerating structure cavity has a bandwidth in kHz range and the 2 MHz offset is far from the resonance of the structure. As Figure \ref{fig-12} shows, the cavity forward and refection signals have extremely similar waveforms, which indicate that the rf power forwarded to the structure is almost fully reflected.

\section{Conclusion}

In this high-power test of the C\(^3\) accelerating structure driven by NG-LLRF, the amplitude and phase stability levels have been evaluated comprehensively at different stages of the test setup. The direct loop-back characterization test of the NG-LLRF prototype demonstrated around 70 fs phase jitters and 0.13 \% of amplitude fluctuation. This is a critical milestone for the new direct rf sampling technique employed to implement the NG-LLRF platform. The rf pulses are directly generated by the DACs and measured by the ADCs without any analogue rf mixing in higher order Nyquist zones, which can significantly simplify the structure of the LLRF system for new accelerator developments and upgrades for existing ones. For instance, the LCLS-II has 35 cryomodules and each of them requires 4 rf inputs and 2 RF outputs for control and monitoring purposes. The RF control system for LCLS-II requires at least 210 independent RF channels with up and down mixers with local oscillator distributed to each of and discrete data converters with conventional LLRF control platforms. With the NG-LLRF platform, all the analog rf mixers can be eliminated, and the integration level is highly improved. Therefore, the NG-LLRF platform can deliver significant power and footprint reductions, and higher adaptability compared with conventional architectures. 

As the up and down conversions of the rf signals are fully realized in digital domain for NG-LLRF, the system can be rapidly configured for different operation modes for rf stations or adapted to accelerators operated at rf frequency lower than 6 GHz by using software-defined methods.The development of the NG-LLRF platform is still in progress. To maximize the adaptability and scalability of the NG-LLRF across applications with different requirements, we are employing modularized approaches to develop the hardware, firmware and software components of the system.

The pulse-to-pulse phase jitter levels have been measured at different stages of the high-power test setup at different power levels. With the peak power at 16.45 MW, the phase jitter measured at the cavity forward coupler is 167.0 fs. The SSA added 82.6 fs phase jitter on top of the direct loopback phase noise, which is the largest among the stages. With the rf signal generation and measurement precision of NG-LLRF, the 150 fs phase jitter level required by C\(^3\) is highly achievable with a real-time feedback control loop.

The high-power test with square rf pulse reveals the full filling and dissipation processes and other reflections between the rf components. On the cavity forward signal, there is a decline after the initial ramp up, which is the result of cross coupling between forward and reflection couplers. For accelerators operating with bunch trains, the requirement in the flatness of the pulse top is more stringent. Therefore, the designs and setup of the couplers to sample the rf signals and the connection waveguides should be tuned to minimize the cross coupling. Digital signal processing techniques, such as the subtraction applied in this case, can also be implemented to further improve the measurement accuracy of rf signals at different stages of the structure. For the tests for rf pulse with phase reversal, the high power short pulses have been successfully generated on the reflection signal. Therefore, the NG-LLRF can be employed to drive and control a SLED structure for pulse compression without any analogue modulator or phase shifter. For the tests of rf pulse with a linear phase sweep in 500 ns, the klystron generates the phase ramp with high precision, but rf power forwarded to the cavity has been fully reflected, as the phase ramp is equivalent of drive the structure off the resonance by approximately 2 MHz.

\section*{Acknowledgment}

The authors express their gratitude to Zenghai Li and  Valery Dolgashev for their insightful discussions, which have significantly contributed to this study. The work of the authors is supported by the U.S. Department of Energy under Contract No. DE-AC02-76SF00515.

\section*{Data Availability Statement}

The data underlying this article will be shared on reasonable request
to the corresponding author.

\nocite{*}
\bibliography{bibliography}

\end{document}